\documentclass[showpacs,preprintnumbers,amsmath,amssymb]{revtex4}

\usepackage{graphicx}
\usepackage{dcolumn}
\usepackage{bm}
\usepackage{url}
\usepackage{epsfig}
\usepackage{multirow}
\usepackage{tensor}
\usepackage{empheq}
\usepackage{amsmath}
\usepackage{color}
\usepackage{leftidx}
\definecolor{myblue}{rgb}{.8, .8, 1}

\def\be{\begin{equation}}
\def\ee{\end{equation}}
\def\ba{\begin{eqnarray}}
\def\ea{\end{eqnarray}}

\newcommand{\fr}[2]{\frac{#1}{#2}}

\def\w{\textrm{w}}

\def\Omo{\Omega_{\rm{m}0}}

\def\Omz{\Omega_{\rm{m}}(z)}
\def\Odez{\Omega_{\rm{de}}(z)}

\def\sigal{\sigma_{\rm{gal}}}
\def\sig8{\sigma_{8}}
\def\sigo8{\sigma_{8}^{0}}

\newcommand{\JCAP}{J.\ Cosmol.\ Astropart.\ Phys.}

\def\ga{\mathrel{\raise.3ex\hbox{$>$\kern-.75em\lower1ex\hbox{$\sim$}}}}
\def\la{\mathrel{\raise.3ex\hbox{$<$\kern-.75em\lower1ex\hbox{$\sim$}}}}

\begin{document}

\title{Measuring bias from unbiased observable}


\author{Seokcheon Lee}
\email[]{skylee@kias.re.kr}
\affiliation{School of Physics, Korea Institute for Advanced Study, Heogiro 85, Seoul 130-722, Korea}

\leftline{KIAS-P14069}


\begin{abstract}
Since Kaiser introduced galaxies as a biased tracer of the underlying total mass field, the linear galaxies bias, $b(z)$ appears ubiquitously both in theoretical calculations and in observational measurements related to galaxy surveys. However, the generic approaches to the galaxy density is a non-local and stochastic function of the underlying dark matter density and it becomes difficult to make the analytic form of $b(z)$. Due to this fact, $b(z)$ is known as a nuisance parameter and the effort has been made to measure bias free observable quantities. We provide the exact and analytic function of $b(z)$ which also can be measured from galaxy surveys using the redshift space distortions parameters, more accurately unbiased observable $\beta \sigma_{\rm{gal}} = f \sigma_8$. We also introduce approximate solutions for $b(z)$ for different gravity theories. One can generalize these approximate solutions to be exact when one solves the exact evolutions for the dark matter density fluctuation of given gravity theories. These analytic solutions for $b(z)$ make it advantage instead of nuisance.

\end{abstract}


\maketitle

\section{linear bias}
\renewcommand{\theequation}{1-\arabic{equation}}
\setcounter{equation}{0}

There have been many models for the evolution of the bias $b(z)$ derived from empirical knowledge \cite{0112162, 14055521}, theory \cite{Fry, 9608004, 9804067}, simulations and from observations which account for the growth and merging of collapsed structure \cite{10013162, 0409314, 11061183}. However, all of these bias fitting forms include the unknown free parameters which need to be fitted with the set of galaxy bias data and simulation. It is shown that an incorrect bias model causes a shift in measured values of cosmological parameters \cite{14055521}. Thus, the accurate modeling to $b(z)$ is prerequisite for the precision cosmology. We obtain the exact linear bias obtained from its definition and show its dependence both on cosmology and on gravity theory. We provide $b(z)$ which can be obtained from both theory and observation. This analytic solution for the bias allows one to use it as a cosmological parameter instead of a nuisance one.

The observed linear galaxy power spectrum using a fiducial model including the effects of bias and the redshift space distortions is given by \cite{0307460}
\be P_{\rm{gal}}^{'}(k',z) = \fr{1}{f_{\parallel}^2 f_{\perp}^2} b^2 P_{\rm{m}}(k,z_0) ( 1 + \beta \mu^2)^2 \Biggl(\fr{g(z)}{g(z_0)}\Biggr)^2 \label{Pgal} \, , \ee
where $f_{\parallel} = \hat{H}(z)/H(z)$ (ratio of the Hubble parameter of the adopted fiducial model, $\hat{H}$ to that of the true model, $H$), $f_{\perp} = D_{\rm{A}}(z)/\hat{D}_{\rm{A}}(z)$ (ratio of the angular diameter distance of the true model, $D_{\rm{A}}$ to that of the adopted fiducial model, $\hat{D}_{\rm{A}}$), $b$ defining the linear bias factor, $P_{m}(k,z_0)$ means the present matter power spectrum, the redshift space distortions (RSD) parameter, $\beta$ is defined as $\beta(k,z) = f(k,z)/b(k,z)$, and $g(z)$ is the linear growth factor of the matter fluctuation, $\delta$ with $z_0$ meaning $z=0$. If one adopts the definition of the linear bias as $b(z) = \sigal(z)/\sig8(z)$, then one obtains $\beta \sigal = f \sig8$. Both $\beta$ and $\sigal$ are obtained from observations, and theories predict $f$ and $\sig8$.

If one takes the derivative of $f(k,z) \sig8(z)$ with respect to $z$, then one obtains ({\it we use $f \sig8$ for $f(k,z) \sig8(z)$ below})
\ba b(k,z) &=& \fr{3}{2} \Omz \sigal(z) \Biggl[ \Bigl(\fr{1}{2} - \fr{3}{2} w \Odez \Bigr) f \sig8 - (1+z) \fr{d(f\sig8)}{dz} \Biggr]^{-1} \nonumber \\ &=& \fr{3}{2} \Omz \sigal(z) \Biggl[ \Bigl(\fr{1}{2} - \fr{3}{2} w \Odez \Bigr) \beta \sigal - (1+z) \fr{d( \beta \sigal)}{dz} \Biggr]^{-1} \label{bkz} \, , \ea where we use $f = d \ln g / d \ln a$, $\sig8(z) = \sigo8 g(z)/g(z_0)$, $\sigal(z)$ denoting the observed fractional rms in galaxy number density, $\Omz = \rho_{\rm{m}}(z)/(\rho_{\rm{m}}(z)+ \rho_{\rm{de}}(z))$, and $\Odez = 1 - \Omz$ (under the assumption of the flat Universe), respectively. One can refer the appendix for detail derivation. All quantities in the second equality of Eq.  (\ref{bkz}) are measurable from galaxy surveys. Both $\sigal$ and $\beta \sigal$ are measured from galaxy surveys \cite{12044725}. Also $\Omo$ can be directly measured from $\sigal$ and $\beta \sigal$ \cite{13076619}. Thus, one can measure the time evolution of bias if there exists enough binned data to measure $\fr{d( \beta \sigal)}{dz} \simeq \fr{\Delta (\beta \sigal)}{\Delta z}$. Future galaxy surveys will provide the sub-percent level accuracy in measuring $\beta \sigal$ \cite{13076619} and will make the accurate measurement of bias possible. This Eq. (\ref{bkz}) holds for any gravity theory because it is derived from its definition.

From the above Eq. (\ref{bkz}), one can understand the theoretical motivation for the formulae of $b(z) \propto g(z)^{-1}$ \cite{0112162, Fry, 10013162}. If one assumes $\sigal$ is constant, then one obtains $b(z) = \fr{\sigal g(z_0)}{\sig8^0} g(z)^{-1} \equiv b_0 g(z)^{-1}$. Thus, the magnitude of $b_0$ is determined by the measured value of $\sigal$ which might depend on luminosity, color, and spectral type of galaxies. However, there is no reason to believe that $\sigal$ is time independent. Thus, we regard $\sigal$ as a time dependent observable in Eq. (\ref{bkz}). In addition, the time evolution of bias is completely determined from observations of $\sigal$ and $\beta \sigal$.

We assume the form of $\sigal(z) = \sigma_{\rm{gal}}^0 (1+z)^{0.1}$ to investigate its behavior where we assign the dependence of bias on galaxy properties into $\sigma_{\rm{gal}}^0$. In this case, the galaxy dependence on $b(k,z)$ is absorbed in $\sigma_{\rm{gal}}^0$ solely. The cosmological dependence on bias is represented by $w$, $\Omz$, and $f \sig8$. Actually, $f \sigma_8$ depends on $w$, $\Omo$, and the underlying gravity theory.

We restrict our consideration for the linear regime and one can solve the sub-horizon solution for the $\delta$ to obtain the growth factor, $g(z)$ for the given model. One can numerically solve this for given models. Even though we just investigate the constant dark energy equation of state $w$CDM, $f(R)$, and DGP model in this {\it Letter}, one can generalize the consideration for the any model by solving $\delta$ numerically.

\subsection{$\w$CDM vs $\Lambda$CDM}
\begin{figure}
\centering
\vspace{1.5cm}
\begin{tabular}{cc}
\epsfig{file=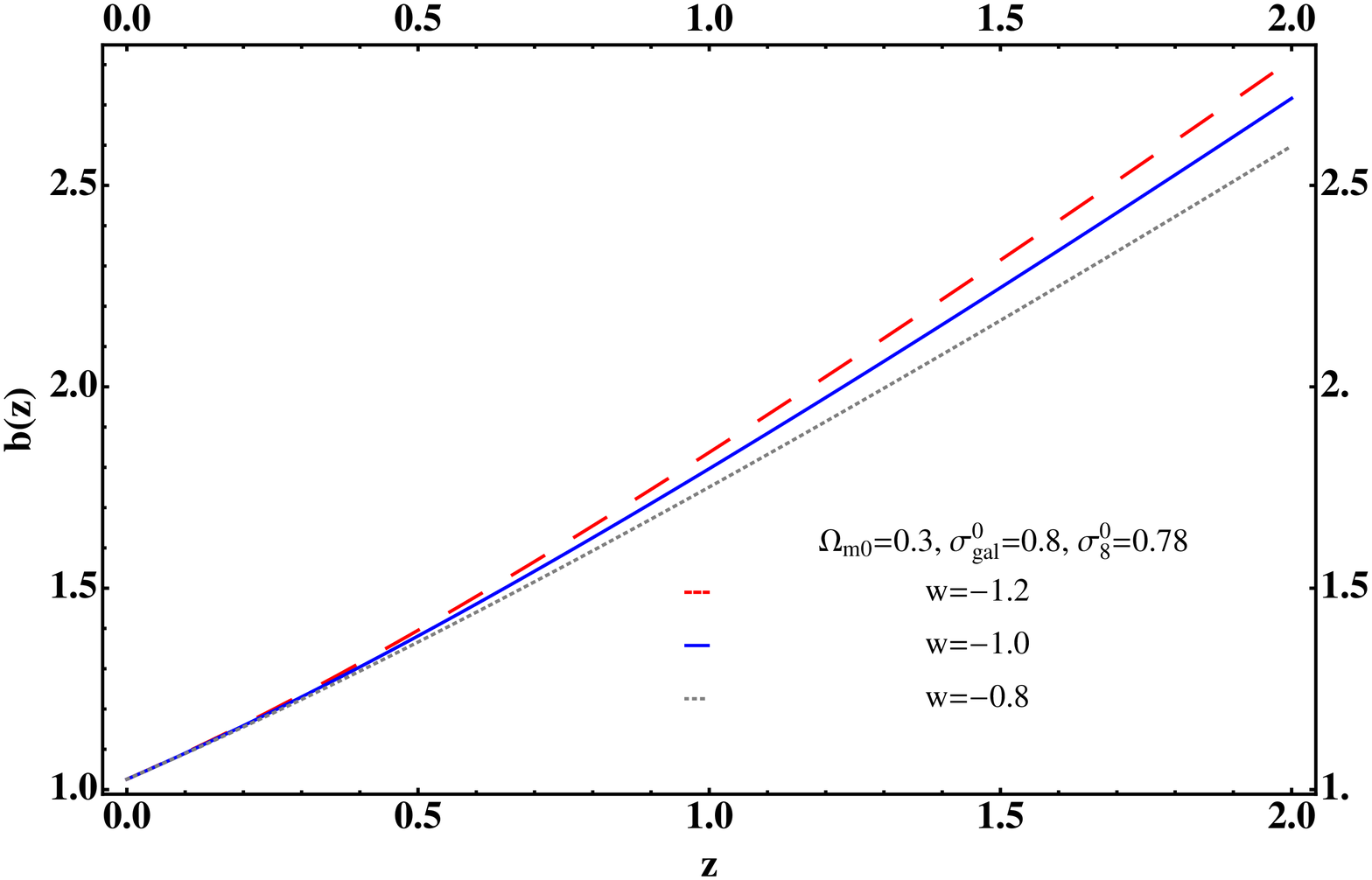,width=0.5\linewidth,clip=} &
\epsfig{file=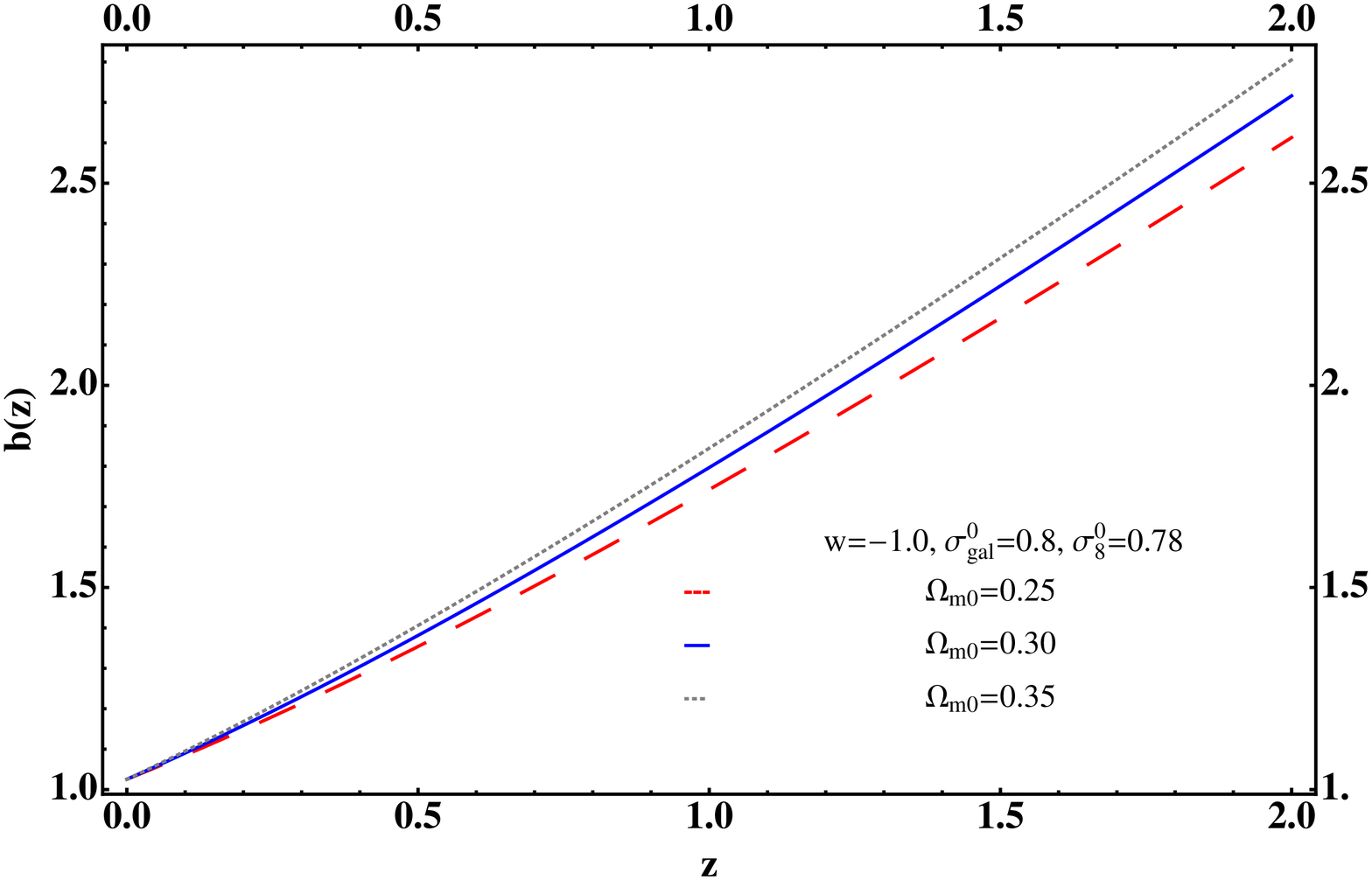,width=0.5\linewidth,clip=} \\
\end{tabular}
\vspace{-0.5cm}
\caption{Time evolution of $b(z)$ for different values of $w$ and $\Omo$ with $\sigma_{\rm{gal}}^0 = 0.8$ and $\sigma_8^0 = 0.78$. {\it Left}) $b(z)$ for $w = $ -1.2 (dashed), -1.0 (solid), and -0.8 (dotted), respectively. {\it Right}) $b(z)$ for $\Omo = $ 0.25 (dotted), 0.30 (solid), and 0.35 (dashed), respectively. } \label{fig1}
\end{figure}

In this subsection, we investigate the evolution of bias for different cosmological parameters ($w$ and $\Omo$) under the General Relativity (GR). For the constant dark energy equation of state, $w$, there exists the known exact analytic solution for the linear growth rate, $g(z)$ \cite{SW, 09072108}. We adopt this solution to show both the cosmology and the astrophysics dependence on $b(z)$. One can generalize the time dependent $w$ by using the numerical solution for the $g(z)$. We depict the dependence of $b(z)$ on $w$ and $\Omo$ in Fig. \ref{fig1}. In the left panle of Fig. \ref{fig1}, we show the evolution of $b(z)$ for different values of $w$ fixed $\Omo = 0.3$, $\sigma_{\rm{gal}}^0 = 0.8$, and $\sigma_8^0 = 0.78$. The dashed, solid, and dotted lines correspond $w =$ -1.2, -1.0, and -0.8, respectively. As $w$ decreases, so does $b(z)$. This is due to the fact that if $w$ increases, then both $g(z)$ and $f(z)$ decrease. The difference of $b(z)$ between models increases, as $z$ increases. The difference between $w = -0.8 (-1.2)$ and $w= -1.0$ is about 4.4 (3.5) \% at $z = 2$.  We also show the $b(z)$ dependence on $\Omo$ for $\Lambda$CDM model in the right panel of Fig. \ref{fig1}. The dashed, solid, and dotted lines correspond $\Omo =$ 0.35, 0.3, and 0.25, respectively for $\Lambda$CDM model. As $\Omo$ increases, so do $g(z)$ and $f(z)$. Thus, $b(z)$ decreases as $\Omo$ increases. The difference between $\Omo = 0.25 (0.35)$ and $\Omo = 0.3$ is about 3.8 (3.2) \% at $z = 2$. Even though we limit our consideration for the constant $w$ with the flat Universe, one can generalize the investigation for the time varying $w$ and the non-flat Universe by solving the sub-horizon equation numerically. Also one can find the time varying $w$ model which produce the same CMB result for the constant $w$ models \cite{14091355}.

\begin{figure}
\centering
\vspace{1.5cm}
\begin{tabular}{cc}
\epsfig{file=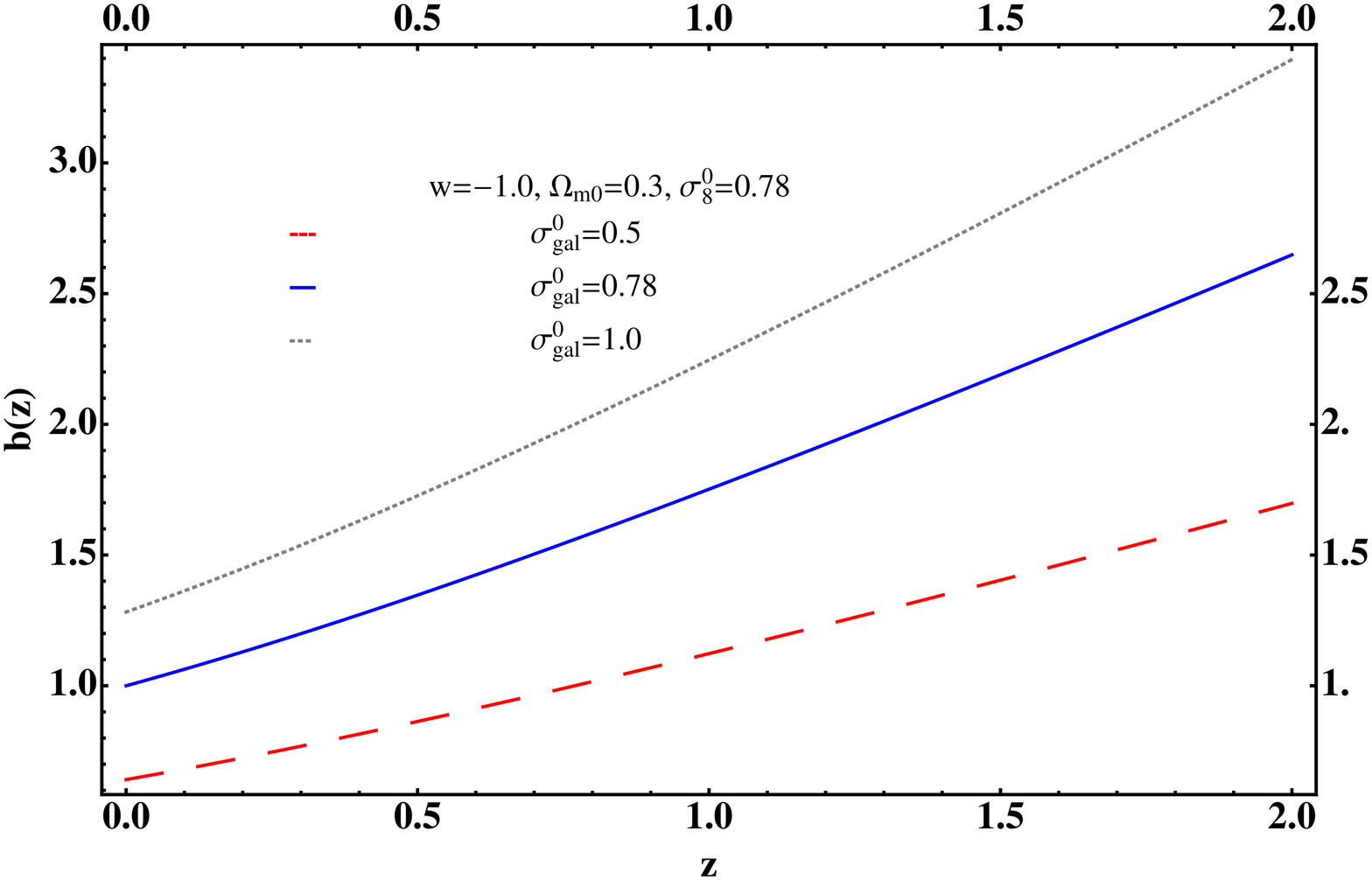,width=0.5\linewidth,clip=} &
\epsfig{file=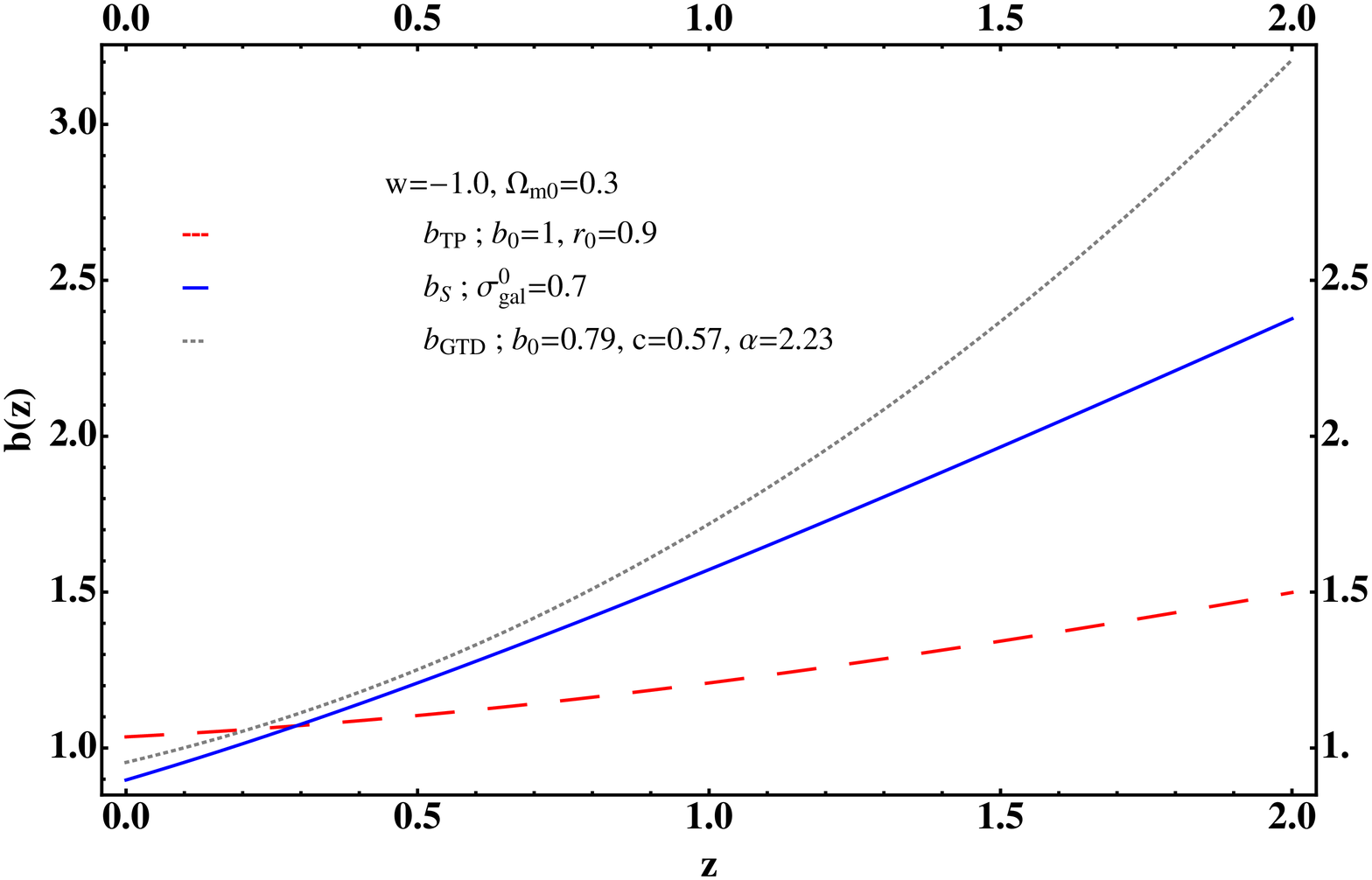,width=0.5\linewidth,clip=} \\
\end{tabular}
\vspace{-0.5cm}
\caption{{\it Left}) Time evolution of $b(z)$ for different values of $\sigma_{\rm{gal}}$ with $w=-1$, $\Omo =0.3$, and $\sigma_8 = 0.78$. $b(z)$ correspond $\sigma_{\rm{gal}} = $ 0.5 (dotted), 0.78 (solid), and 1.0 (dashed), respectively. {\it Right}) $b(z)$ for different models. $b_{\rm{TP}}$ (dashed), $b_{\rm{S}}$ (solid), and $b_{\rm{GTD}}$ (dotted), respectively.} \label{fig2}
\end{figure}

The $b(z)$ dependence on $\sigma_{\rm{gal}}^0$ is also shown in the left panel of Fig. \ref{fig2}. If one adopts $\sigma_{\rm{gal}}^0 < \sigma_8^0$, then the present value of $b(z)$ is smaller than unity. This can explain IRAS result that galaxies are less clustered than the dark matter distribution \cite{Saunders} even though the current measurement shows the correlated result \cite{13090382}. Also if $\sigma_{\rm{gal}}^0 > \sigma_8^0$, then the present value of $b(z)$ is larger than the unity. The dotted, solid, and dashed lines correspond $\sigma_{\rm{gal}}^0 = 1.0$, 0.78, and 0.5, respectively. We also compare the fitting formula of $b(z)$ with ours for $\Lambda$CDM model in the right panel of Fig \ref{fig2},
\ba b_{\rm{GTD}} &=& c + (b_0 - c) g(z)^{-\alpha} \label{bGTD} \\
b_{\rm{TP}} &=& \sqrt{(1-g(z))^2 - 2(1-g(z)) b_0 r_0 + b_0^2} \Bigl / g(z) \label{bTP} \, . \ea
The dotted, solid, and dashed lines correspond GTD \cite{14055521}, ours (S), and TP \cite{9804067}, respectively. Ours are less steep than GTD and steeper than TP. Also GTD always predict the lower than unity at the present epoch, but our can be any value depends on $\sigma_{\rm{gal}}^0$. The difference between $b_{\rm{GTD}} (b_{\rm{TP}})$ and $b_{\rm{S}}$ is about 35 (37) \% at $z=2$. One can find more fitting forms of bias \cite{14055521}.

\subsection{Constrains on b(z) and paramters}

\begin{figure}
\centering
\vspace{1.5cm}
\begin{tabular}{ccc}
\epsfig{file=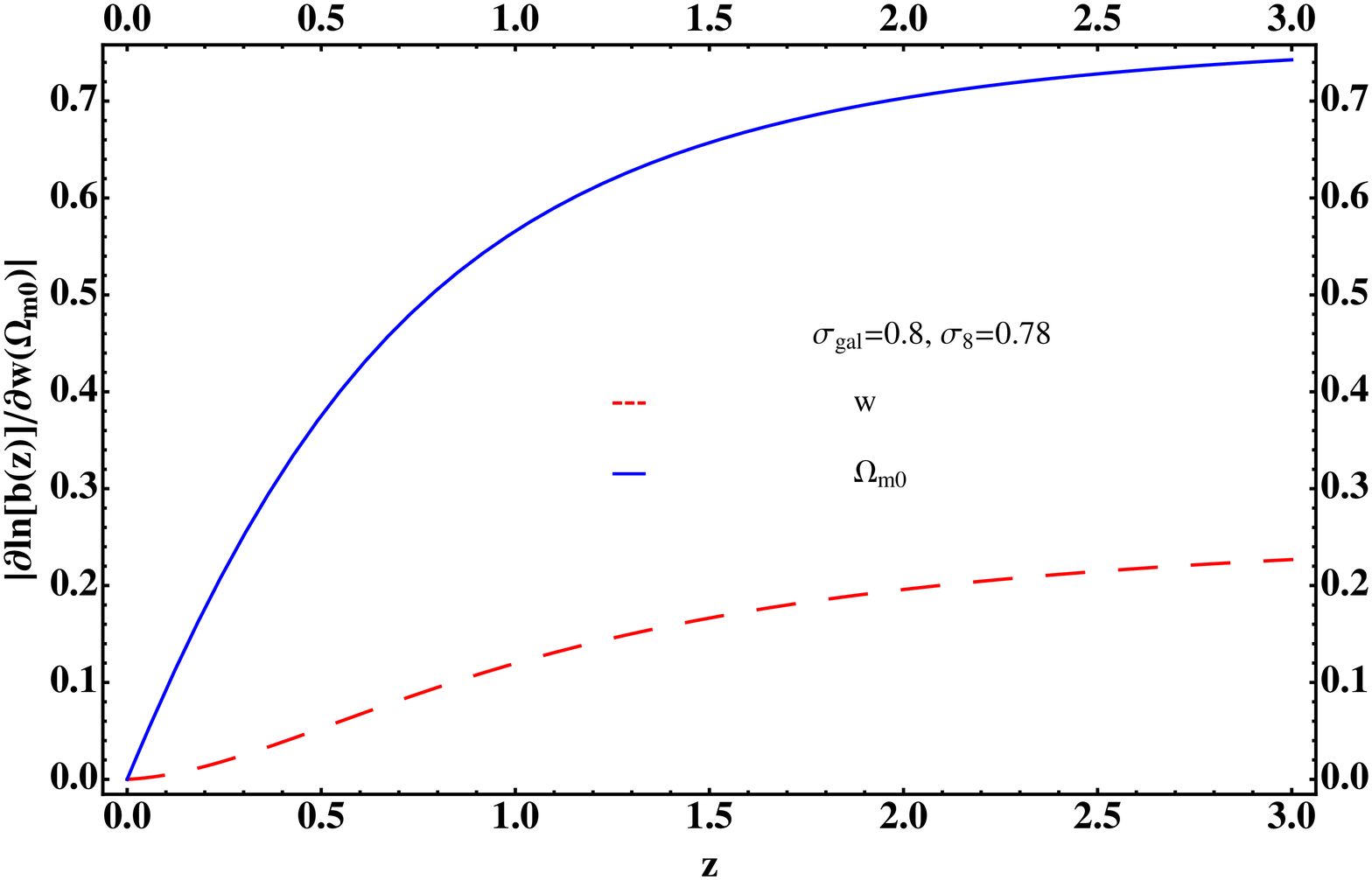,width=0.44\linewidth,clip=} &
\epsfig{file=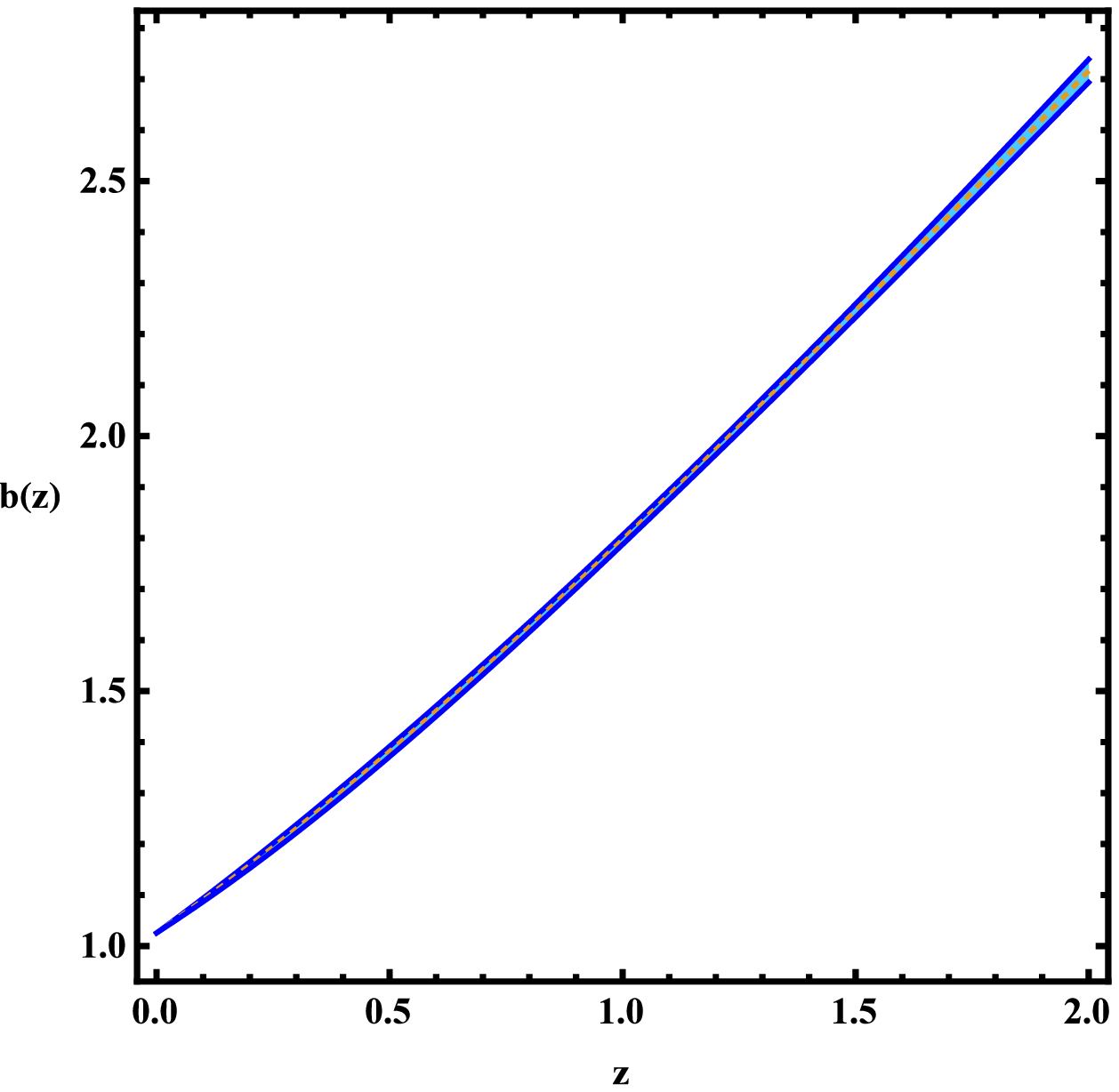,width=0.27\linewidth,clip=} &
\epsfig{file=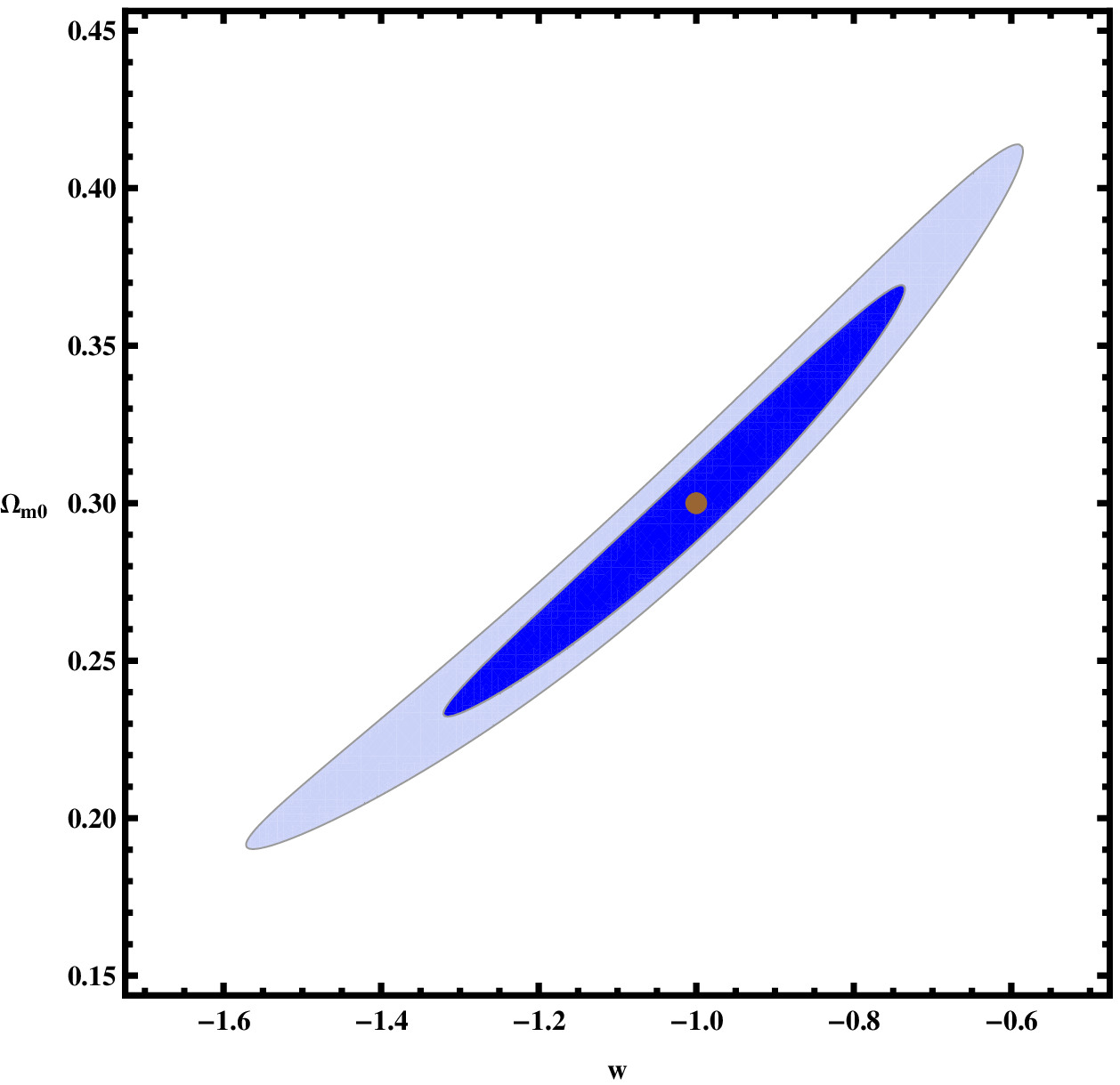,width=0.27\linewidth,clip=}
\end{tabular}
\vspace{-0.5cm}
\caption{{\it Left}) The sensitivity of the $b(z)$ to the cosmological parameters, $w$ and $\Omo$. The thick solid line and the dashed line correspond $\Omo$ and $w$, respectively. {\it Middle}) The variation of the $b(z)$ over redshift. The dashed line are the best-fit and the shaded shadow contour represents the 68 \% confidence level around the best fit. {\it Right}) The 1-$\sigma$ (inner ellipsoid) and 2-$\sigma$ (outer one) confidence contours in the $w$-$\Omo$ plane for the corresponding $b(z)$. } \label{fig3}
\end{figure}

One can understand the dependence of bias on parameters by using Fisher matrix analysis. First, we investigate the sensitivity of $b(z)$ to the parameters, $w$ and $\Omo$. The sensitivity of its estimation of parameter values depends on the derivatives of the $b(z)$ with respect to the parameter, $\partial b(z) / \partial w (\Omo)$, and the precision with which the observations can be made. This is shown in the left panel of Fig.\ref{fig3}. Sensitivities on both $w$ and $\Omo$ grow as $z$ increases. Sensitivities on $w$ and $\Omo$ have the opposite sign ($\partial b(z) / \partial w$ is negative and $\partial b(z) / \partial \Omo$ is positive) and it means the error ellipse in the $w-\Omo$ contour plot have the opposite orientation as for the supernova distance case. Making $w$ larger can be acted by making $\Omo$ larger. This fact provides the promise of complementarity with supernova. The larger the absolute magnitude of the derivative at a particle redshift, the stronger constrain on the bias.   If future galaxy redshift surveys constrain $\sigma_{\rm{gal}}$ and $\beta \sigma_{\rm{gal}}$ at percent levels, then $b(z)$ is also constrained at the same level \cite{13076619}. In this forecast, we adopt measurement errors on $b(z)$ as 2 \%. Then the time evolution of the bias $b(z)$ is obtained as in the middle panel of Fig.\ref{fig3}. The central dotted line is the $b(z)$ from the best fit value and the shadow contours represent the 1-$\sigma$ confidence level around the best fit. As one expects, $b(z)$ is tightly constrained. In the right panel of the same figure, we show the 1-$\sigma$ (inner ellipsoid) and 2-$\sigma$ (outer one) confidence contours in the $w-\Omo$ plane. As we show in the sensitivity figure, $\Omo$ has the stronger constrain than that of $w$. 1-$\sigma$ for $w$ ($\Omo$) is 0.190 (0.045).

\subsection{General Relativity vs Modified Gravity}
In this subsection, we investigate the time evolution of bias for different gravity theories. One can use the approximate solution for the growth rate for the different gravity model specified by $\gamma_i$,
\be f_i \equiv \fr{d \ln g_i}{d \ln a} \approx \Omega_{\rm{m}}^{\gamma_i}(z) \label{fapp} \ee
One can integrate the above equation to obtain $g_i(z)/g_i(z_0) = \exp \Bigl[\int_{-\ln [1+z]}^{0} \Omega_{\rm{m}}^{\gamma_i}(z') d \ln (1+z') \Bigr]$ where we use $\gamma_{\rm{DGP}} \simeq 11/16$ \cite{DGP}, $\gamma_{\rm{GR}} \simeq 0.555$ \cite{09061643}, and $\gamma_{\rm{f(R)}} \simeq 0.41$ \cite{fR}. We assume that background evolutions of all model mimic the that of $\Lambda$CDM.

\begin{figure}
\centering
\vspace{1.5cm}
\begin{tabular}{cc}
\epsfig{file=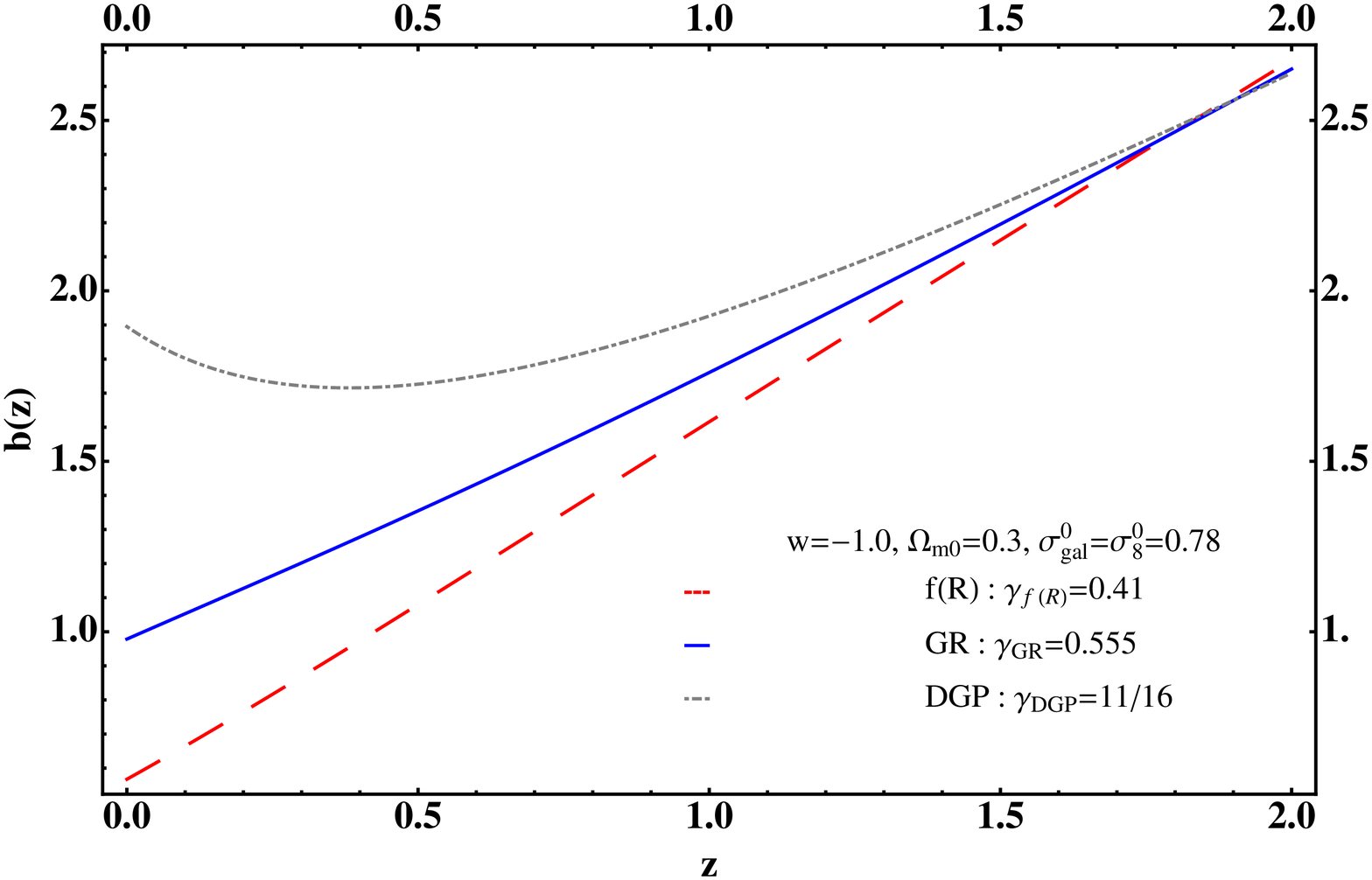,width=0.5\linewidth,clip=} &
\epsfig{file=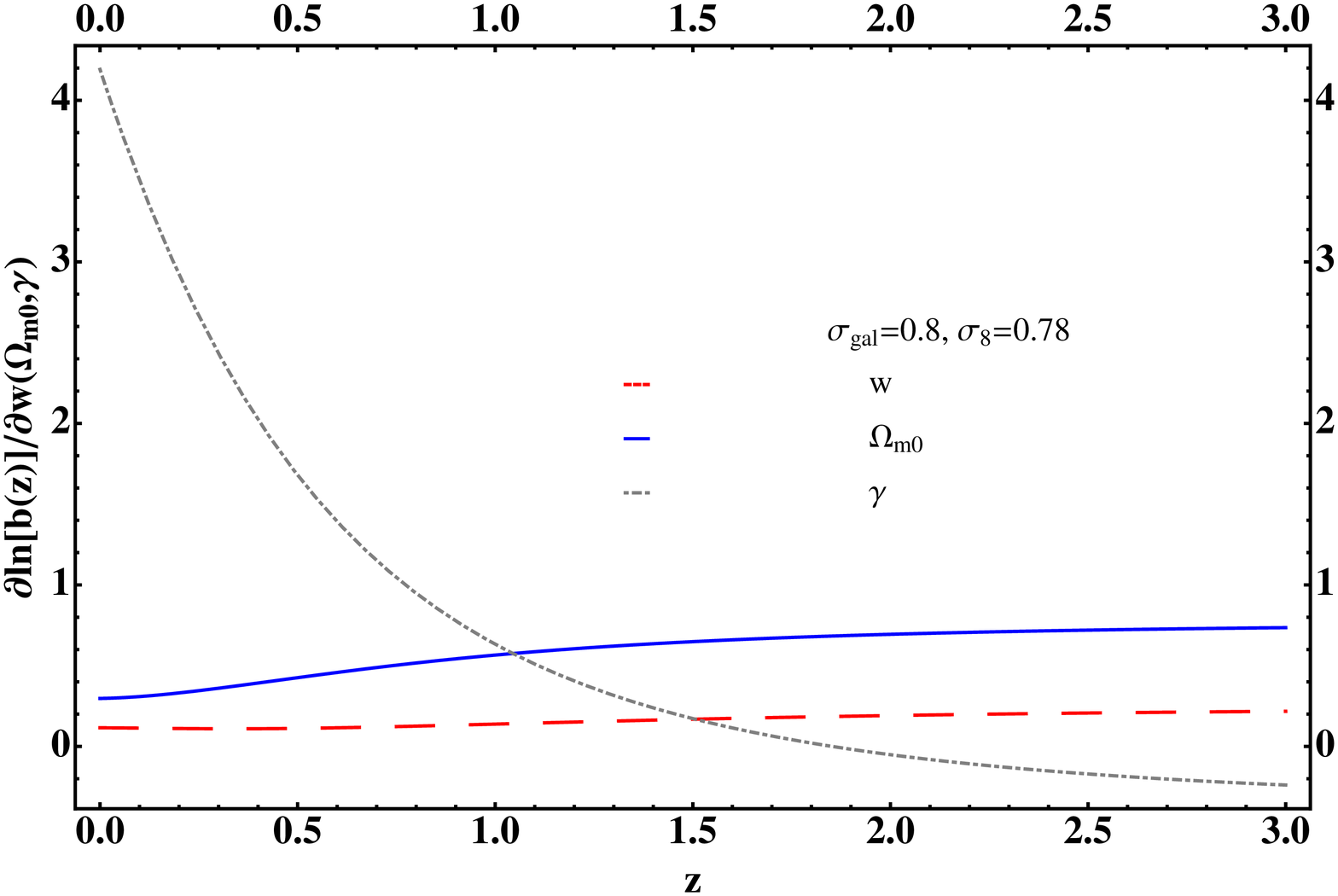,width=0.5\linewidth,clip=} \\
\end{tabular}
\vspace{-0.5cm}\caption{ {\it Left}) $b(z)$ for different gravity theories with assuming the same background evolution as that of GR with $w =-1.0$, $\Omo = 0.3$. We also adopt $\sigma_{\rm{gal}}^0 = \sigma_8^0 = 0.78$. Lines correspond $f(R)$ (dashed), GR (solid), and DGP (dotted), respectively. {\it Right}) The sensitivity of the $b(z)$ to the cosmological parameters, $w$, $\Omo$, and $\gamma$. The solid (dashed, dot-dashed) line correspond $\Omo$ ($w$, $\gamma$). }
\label{fig4}
\end{figure}

Interesting feature for different gravity theories comes from deviations of $b(z)$s at low redshifts. DGP shows the fastest growing of the growth factor among models to produce the largest $b(z)$ at the present epoch. f(R) gravity produces the smallest growth factor to give the smallest $b(z)$ at $z_0$. The difference of bias between f(R) (DGP) and GR is about 42 (94) \% at $z_0$. These are shown in the left panel Fig. \ref{fig4}. The dashed, solid, and dotted lines correspond $b_{\rm{f(R)}}$, $b_{\rm{S}}$, and $b_{\rm{DGP}}$, respectively. The differences of $b(z)$s between models decrease as $z$ increases. This is due to the fact that we adopt the different gravity theories converge to normal $\Lambda$CDM background around $z > 2$. Thus, one should be careful for the bias when one adopts different gravity theories in galaxy surveys. We also investigate the sensitivity of the $b(z)$ to the cosmological parameters, $w$, $\Omo$, and $\gamma$. As one already see in the first panel of Fig.\ref{fig4}, the $\gamma$ provides the strongest constrain on $b(z)$ around $z < 1$. Thus, the measurement of $b(z)$ can be used to break the known degeneracy between the equation of state $w$ and the growth index parameter $\gamma$ due to the evolution of $\Omz$ \cite{09103834}.

\section{Conclusions}
\setcounter{equation}{0}
We obtain the exact analytic solution for the linear bias. This solution can investigate both cosmological and astrophysical dependence on bias without any ambiguity. From this solution, one can exactly estimate the time evolution of bias for different models. The different gravity theories provide the different bias. Thus, this provides the consistent check for the cosmological dependence on the measured galaxy power spectrum for the given model. This solution can be generalized to many models including the modified gravity theories and the massive neutrino dark matter model by replacing the approximate solution used in this {\it Letter} with the exact sub-horizon solutions for corresponding models. These cases are under investigation \cite{SL}. This theoretical form of bias can be measured from measurements of $\sigal$ and $\beta \sigal$ from galaxy surveys if we achieve enough binned data. Also a known degeneracy between the equation of state $w$ and the growth index parameter $\gamma$ due to the evolution of $\Omz$ can be broken due to this exact form of bias and can be used to distinguish the dark energy from the modified gravity.

\section*{Acknowledgments}
We would like to thank Xiao-Dong Li and Hang Bae Kim for useful discussion. This work were carried out using computing resources of KIAS Center for Advanced Computation. We also thank for the hospitality at APCTP during the program TRP.

\renewcommand{\theequation}{A-\arabic{equation}}
\setcounter{equation}{0}  
\section*{APPENDIX}  
One takes the derivative of $f(a) \sigma_8(a)$ using their definitions, $f(a) = d \ln g / d \ln a$ and $\sigma_8(a) = g(z)/g(z_0) \sigma_8^0$ to obtain
\be \fr{d (f(z) \sigma_8(z))}{dz} =  \fr{1}{(1+z)} \Biggl[ \Bigl(\fr{1}{2} - \fr{3}{2} w \Omega_{\rm{de}}(z) \Bigr) f(z) \sigma_8(z) - \fr{3}{2} \Omz \sigma_8(z) \Biggr] \label{dfsig8dz} \, , \ee where we use the sub-horizon scale equation for the growth factor $g(z)$, $\ddot{g} + 2H \dot{g} - 4\pi G \rho_{m} g = 0$ where dot means the derivative with respect to the cosmic time $t$. Thus, one obtains an interesting relation between $\sigma_8$ and $f \sigma_8$,
\ba \sigma_8(z) &=&  \fr{2}{3 \Omz} \Biggl[ \Bigl(\fr{1}{2} - \fr{3}{2} w \Odez \Bigr) f \sig8 - (1+z) \fr{d(f\sig8)}{dz} \Biggr] \nonumber \\
&=&  \fr{2}{3 \Omz} \Biggl[ \Bigl(\fr{1}{2} - \fr{3}{2} w \Odez \Bigr) \beta \sigal - (1+z) \fr{d(\beta \sigal)}{dz} \Biggr] \label{sigma8} \, , \ea where we explicitly express the $\sig8(z)$ using the observable quantity $\beta \sigal$ in the second equality. Thus, if one achieves enough binned data for $\beta \sigal$, then one can measure $\sig8$ at each epoch. For example, the present value of $\sig8$ is given by
\ba \sigma_8^0 &=& \fr{2}{3 \Omz} \Biggl[ \Bigl(\fr{1}{2} - \fr{3}{2} w (1-\Omo) \Bigr) \beta \sigal - \fr{d(\beta \sigal)}{dz} \Biggr] \Biggl( \fr{g(z)}{g(z_0)} \Biggr)^{-1} \nonumber \\
&=& \fr{2}{3 \Omo} \Biggl[ \Bigl(\fr{1}{2} - \fr{3}{2} w (1-\Omo) \Bigr) \beta_0 \sigma_{\rm{gal}}^0 - \fr{d(\beta \sigal)}{dz} \Bigl |_{z_0} \Biggr]  \label{sig80} \, . \ea The value of $\sigma_8^0$ derived from the CMB depends on the primordial amplitude, $A_s$ and the spectral index, $n_s$. However, the right hand side of Eq. (\ref{sig80}) depends only on the background evolution parameters, $w$ and $\Omo$. Thus, one can the constraint $A_s$ and $n_s$ from the RSD measurement. $\sigma_8$ and $\Omo$ are degenerated in galaxy surveys, but one can break this from the above Eq. (\ref{sig80}). If one adopts the definition of linear bias $b(z) = \sigma_{\rm{gal}}(z) / \sigma_8(z)$, then one obtains $b(z)$ from the above Eq. (\ref{dfsig8dz})
\be b^{-1}(z) = \fr{2}{3 \Omz \sigma_{\rm{gal}}(z)} \Biggl[ \Bigl(\fr{1}{2} - \fr{3}{2} w \Odez \Bigr) f \sig8 - (1+z) \fr{d(f\sig8)}{dz} \Biggr] \label{binv} \, . \ee Thus, one obtains the exact analytic solution for $b(z)$ given by Eq. (\ref{bkz}). One can generalize $b(z)$ as $b(k,z)$ if one substitute $g(z)$ with $g(k,z)$ even for sub-horizon scales. For example, if one considers $f(R)$ model or the massive neutrino model, then one can obtain $g(k,z)$ inside horizon scales at linear regime \cite{SL}.

\end{document}